# Surface nucleation of the paraelectric phase in ferroelectric BaTiO$_3$: Atomic scale mapping

*Maya Barzilay,*[1,2] *Hemaprabha Elangovan*[1,2] *and Yachin Ivry*[1,2,*]

[1]Department of Materials Science and Engineering, Technion – Israel Institute of Technology, Haifa 3200003, Israel.

[2]Solid State Institute, Technion – Israel Institute of Technology, Haifa 3200003, Israel.

[*]Correspondence to: ivry@technion.ac.il.

## Abstract

In ferroelectricity, atomic-scale dipole moments interact collectively to produce strong electro-mechanical coupling and switchable macroscopic polarization. Hence, the functionality of ferroelectrics emerges at a solid-solid phase transformation that is accompanied by a sudden disappearance of an inversion symmetry. Much effort has been put to understand the ferroelectric transition at the polarization length scale. Nevertheless, the dipole-moment origin of ferroelectricity has remained elusive. Here, we used variable-temperature high-resolution transmission electron microscopy to reveal the dipole-moment dynamics during the ferroelectric-to-paraelectric transition. We show that the transition occurs when paraelectric nuclei of the size of a couple of unit cells emerge near the surface. Upon heating, the cubic phase sidewalk grows towards the bulk. We quantified the nucleation barrier and show dominancy of mechanical interactions, helping us demonstrate similarities to predictions of domain nucleation during electric field switching. Our work motivates dynamic atomic-scale characterizations of solid-solid transitions in other materials.

## Introduction

Micro-scale realization of phase transitions that are accompanied by symmetry change is a longstanding goal of scientists from a broad range of disciplines, spanning condensed matter physics,[1,2] materials science,[3] and geology,[4] as well as biology,[5] and even high-energy physics,[6] and sociology.[7,8] Ferroelectrics are functional materials, in which the functionality is coupled to the crystal structure and hence they serve as a convenient platform for understanding phase transition processes.[9–12] Ferroelectrics possess switchable polarization and are therefore vital, e.g. for low power switching devices,[13] wireless filters,[14] biomedical imaging and sensing applications[15–17] and energy harvesting.[18] Ferroelectricity arises when atomic-scale dipole moments organize to form macroscopic electric polarization. These dipoles stem from the lack of centrosymmetry of the crystal phase. Ferroelectricity thus vanishes when a material undergoes a symmetry change from a non-centrosymmetric to a centrosymmetric structure. Such a ferroelectric-to-paraelectric transition typically occurs upon heating the material above a certain Curie temperature, $T_C$.

$BaTiO_3$ is an established lead-free material for ferroelectric applications,[19] which serves as a prototype for exploring ferroic transitions in perovskite materials as well as for general phase transitions that are accompanied by symmetry change. Much effort has been put to realize the ferroelectric-to-paraelectric transition in ferroelectrics both theoretically and experimentally, including in $BaTiO_3$.[12,20–22] Yet, the exact mechanism at which the atomic-scale dipole moments organize during the ferroic transformation has remained unknown, mainly because of the accompanying experimental difficulties. Specifically, the dipole-moment and unit-cell changes in perovskite during the transformation from tetragonal to cubic are only of a few picometers in size, challenging the limits of today's microscopy.

Dipole-moment mapping of static polarization states have been obtained in various ferroelectrics, mainly with high-resolution transmission electron microscopy (HRTEM), examples include mainly $PbTiO_3$ and $BiFeO_3$ lamellae, both materials have relatively large ion displacement.[23–25] Atomic-scale characterization has also been performed for nano-particle geometries of these materials as well as for $BaTiO_3$.[26] For fine grains, size effects suppress the ferroic transition, which is replaced by either complex multi-phase stability over the entire temperature range[27] or by complete suppression of the ferroelectric phase,[28] preventing us from realizing the phase-transition dynamics.

A major hurdle in characterizing the ferroic transition is the short time scale of the process in comparison to contemporary atomic-scale mapping techniques. Dipole dynamics is important not only for phase transitions but also for technological applications. Several studies have reported polarization domain dynamics under electric fields, mainly by means of piezoresponse force microscopy (PFM),[29–31] and even with electron microscopy.[32] However, the atomic-scale dipole-moment dynamics during either domain switching or at the ferroic transition have remained elusive. Likewise, despite the importance of the nucleation or growth mechanism, especially at the unit-cell scale, to the onset of ferroelectricity, most available theories discuss this process in the context of domain switching under electric field and do not address the ferroic transition.[10,33] Thus, current understanding of the atomic-scale dynamics during ferroic transitions rely on indirect studies that discuss electric-field switching rather than temperature variations, encumbering realization of the origin of ferroelectricity.

Recently, domain dynamics were reported by means of direct observation with PFM during the orthogonal-to-tetragonal phase transition in a single-crystal $BaTiO_3$, allowing mesoscopic-scale realization of this transition.[34] To overcome the discrepancy between transition timescale and the experimental imaging setup timescale, the temperature was varied very slow (~1 K/hr) with high temperature stability (noise level ± 0.015 K), allowing real-time polarization mapping during the phase transformation.[34] However, such a temperature control is not yet feasible for atomic-scale mapping. Hence, alternative methods must be sought. Here, we expanded the temperature range of the ferroic transformation rather than slowing down the temperature variation. Figure 1a describes the phase stability map schematically in ferroelectrics as a function of crystal size. The schematics show that for large crystals, the material transfers abruptly from tetragonal to cubic ($r_L$ at $T_{C_0}$, see Figure 1b), whereas in very fine structures, there is no clear distinction between individual tetragonal and cubic phases ($r_{VF}$, Figure 1b). Typically, such small nano particles do not allow domain formation[28] or demonstrate merely phase mixtures and not individual phase states.[27] Yet, at an intermediate crystal size, $r_0$ ($r_{VF} < r_0 < r_C$, where $r_C$ is the smallest size of $r_L$), the transformation between a complete tetragonal phase to a complete cubic structure is stretched so that a cubic-tetragonal coexistence regime appears (between $T_{C_1}$ and $T_{C_2}$, see Figure 1c), allowing us to observe the phase-transition dynamics and resolve the atomic scale emergence of the ferroic transition.

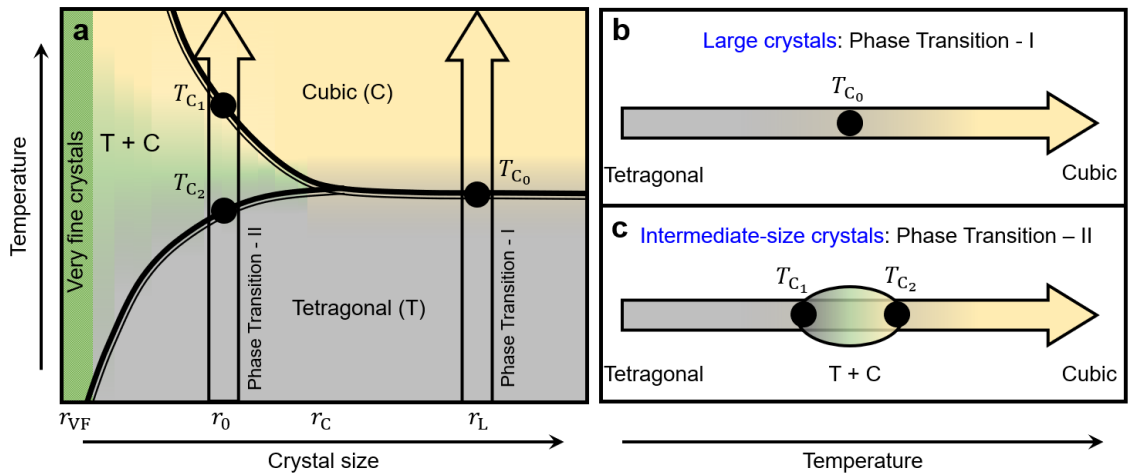

**Figure 1| Schematic phase diagram of BaTiO$_3$ crystals with various sizes.** Schematics of the phase stability map as a function of crystal size (inspired by Y. Li et al.[27] and T. Hoshina et al.[28]). (**a**) Temperature – crystallographic-structure map of the cubic and tetragonal phases in BaTiO$_3$. (**b**) Temperature-driven transition for large crystals (I), showing the abrupt transformation from tetragonal to cubic symmetry at $T_{C_0}$. (**c**) Temperature-driven transition for intermediate-size crystals (II), demonstrating a tetragonal-cubic coexistence region between $T_{C_1}$ and $T_{C_2}$, while at lower (higher) temperatures, a single tetragonal (cubic) phase is dominant.

# Experimental

50-nm BaTiO$_3$ single crystals were dispersed in ethanol and sprayed on the TEM grids by employing pressurized N$_2$ gas (see Reference [35] for details regarding the sample preparation and complementary characterization). In-situ heating was performed with a DENS heating system, in which the grid is embedded in a nano-chip that comprises a temperature controller and a dedicated heating holder is used. We used 300 keV accelerating voltage electron beam in a double-corrected Titan Themis G2 300 (FEI/Thermo Fisher) with sub-angstrom resolution. No manipulation was performed to the acquired images, while the image analyses were performed using Digital Micrograph software (in cases where a closer look was required to highlight to effects, further explanations and larger-scale original images are given in the SI).

X-ray diffraction (XRD) characterization was performed using a Rigaku SmartLab diffractometer with a Cu electrode ($\lambda = 0.15406$ nm) by employing $\theta - 2\theta$ method of measurement. High-temperature XRD measurements were performed using the dedicated heating sample stage.

We chose 50-nm crystal size because on the one hand, the material is large enough to be considered single-crystal and demonstrates distinguishable individual tetragonal and cubic phases above and below the transition, respectively.[36] Yet, on the other hand, such crystals are thin enough to allow non-destructive TEM characterization as well as for exhibiting a tetragonal-cubic coexistence regime (see Figure 1c). Likewise, the small crystal size assures small heat capacity, avoiding any meaningful temperature gradients during the experiments.

## Results

To identify the Curie temperature as well as to demonstrate the existence of a single dominant tetragonal (cubic) phase below (above) the transition, macroscopically, we first performed variable-temperature XRD characterization. Figure 2a shows a split between the (002) and (200) planes in the XRD data at 308 K (red circles). This peak splitting vanishes at 393 K (black squares in Fig. 2a), demonstrating the stability of the respective dominant individual tetragonal and cubic phases.

In addition to the macroscopic XRD characterization, we mapped the dipole moments below and above the transition at the atomic scale. Figure 2b and 2c show the tetragonal dominancy at 303 K (slight heating was applied to assure temperature stability) from [100] and [110] zone axes. The Ti ion off-center displacement is highlighted in these images. Looking at the histogram of the Ti ion displacement from Figure 2b (see Figure SI-1), we found that the dipole moments are 22±5 pm ion displacement. The high confidence in dipole-moment characterization allowed us to map the dipole moment and tetragonality also in multi-domain structures. Figure 2d shows such a structure of in-plane $a_1$-$a_2$ 90° domains (note that this domain wall in the free crystalline involves polarization rotation, which suggests a non-Ising domain wall).[37,38]

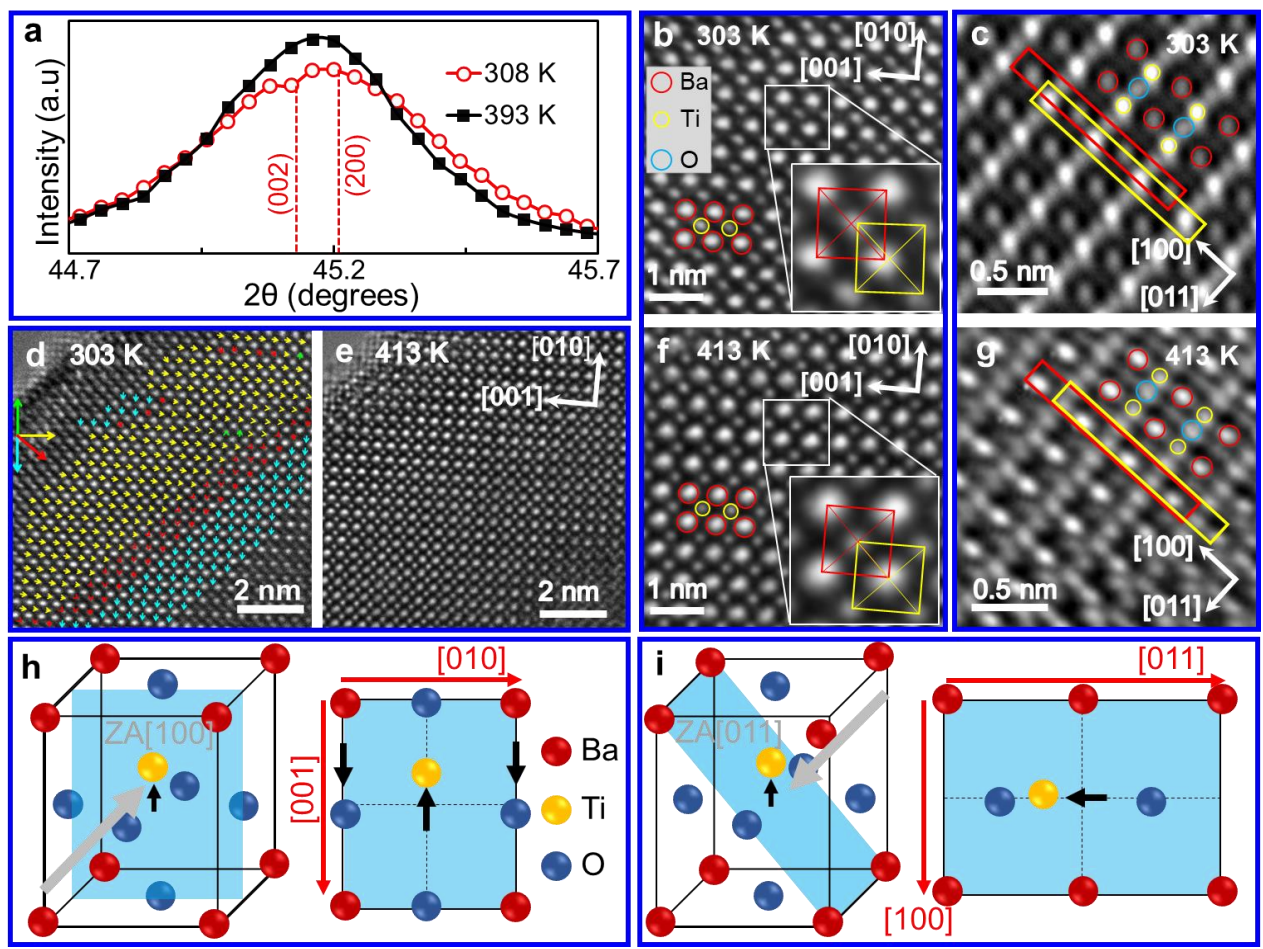

**Figure 2| Single tetragonal and cubic phase dominancy below and above the ferroic transition, respectively, in intermediate-size BaTiO₃ crystals.** (**a**) Macroscopic XRD profiling at 308 K (red circles) and 393 K (black squares), showing the tetragonal and cubic dominancy at temperatures lower and higher than the Curie temperature, respectively. Here, the tetragonal lattice parameters are $c_T = 4.0187$ Å, $a_T = 4.0086$ Å, while the cubic lattice constant is $a_C = 4.0120$ Å. (**b-c**) Atomic-scale mapping (TEM micrographs) of dipole moments within single ferroelectric domains and (**d**) a multi-domain structure in BaTiO₃ crystallites at 303 K, showing tetragonal dominancy from different zone axes. (**e-g**) In-situ heating to 413 K shows cubic symmetry in these areas (blue frames help couple low and high temperature scans from similar areas). (**h**) Schematics of the Ti ion displacement in the tetragonal BaTiO₃ crystal structure and the projections on [001] and (**i**) [011] zone axes to help analyze the dipole-moment mapping in (b-d), zone axes (ZA) are designated by gray arrows.

Next, similarly to demonstrating the tetragonal dominancy below the transition, we wanted to show that above the transition (>393 K, Fig. 2a), the crystal symmetry changes to cubic. Figures 2e-g supply a direct evidence that at 413 K (above $T_C$), all multi-domain (Figure 2e) and single-domain (Figures 2f-g) structures already become a homogeneous cubic phase, as expected.

Once we confirmed the existence of distinguishable individual phases below and above the transition, we aimed at characterizing the transition dynamics. We chose a single-domain structure and performed real-time imaging during in-situ gradual temperature increment from 373 K to 383 K. This heating process last 50 sec, and we assume a linear temperature change over this time.

At 373 K (Figure 3a), 20 K below $T_C$, the entire area was tetragonal. The transition to cubic began already at 375 K (Figure 3b), where a nucleus with a size of about ~1 nm emerged near the crystal surface. Further heating to 377 K revealed the appearance of additional similar nucleation sites (Figure 3c). Increasing the temperature to 379 K and above resulted in merging of these nuclei and transitioning from nucleation to side-walk growth (Figures 3d-f). The sidewalk progressed towards the [011] direction in these images.

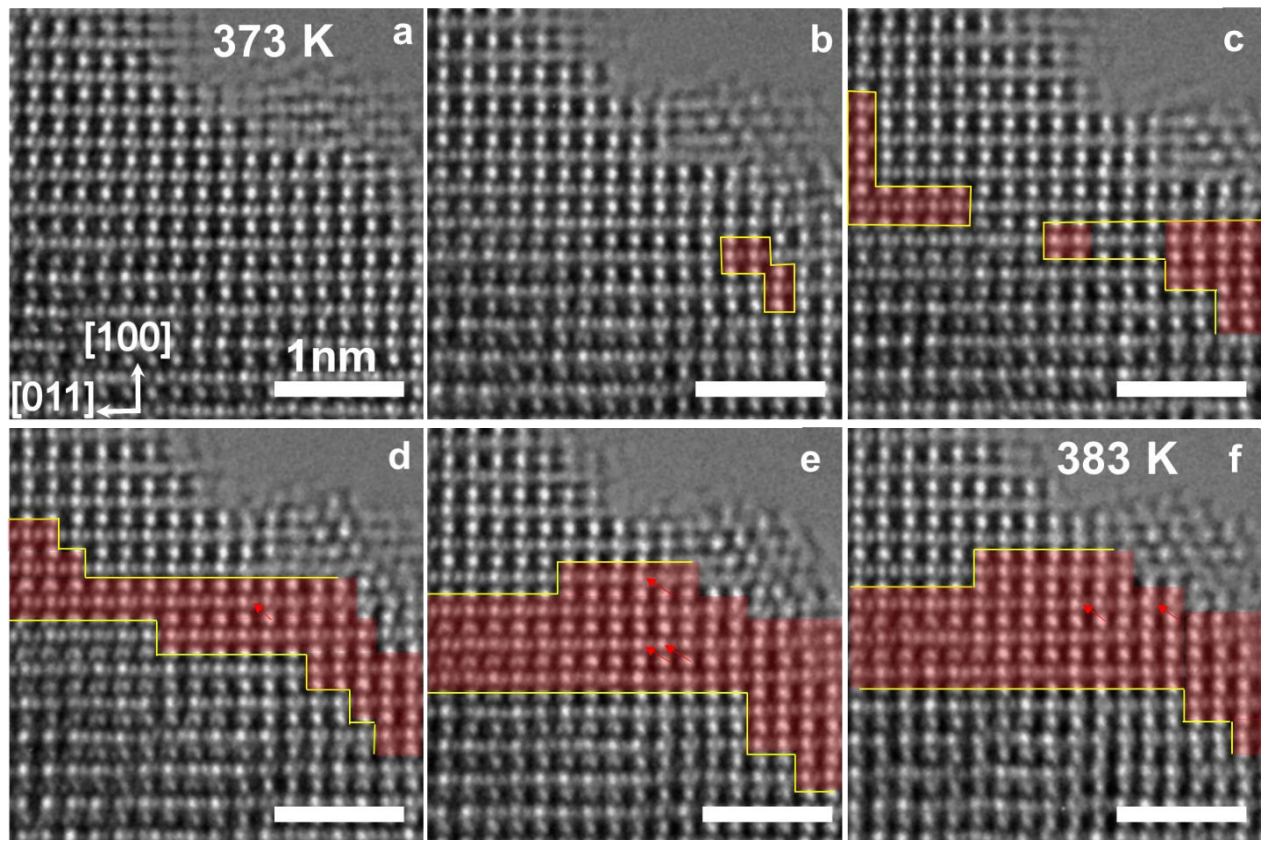

**Figure 3| Temperature-driven nucleation and growth of a paraelectric phase in ferroelectric BaTiO₃.** Atomic-scale mapping (HRTEM) of the tetragonal and cubic unit cells during in-situ heating reveals (**a**) homogeneous tetragonal structure at 373 K; (**b**) emergence of unit-cell size nucleation cubic sites near the surface between 375 K and (**c**) 377 K, (**d**) followed by growth of the cubic phase at 379 K, (**e**) 381 K and (**f**) 383 K (above this temperature, roughly 50% of the unit cells undergone symmetry change, hindering clear distinction between the two phases with the TEM technique that averages out atomic location over the entire crystal thickness).

To characterize the transition dynamics, we quantified the change in cubic/tetragonal unit-cell concentration as a function of temperature (Figure 4). Note that at 383 K (Figure 3f), about half of the material already transferred to cubic, so that confident quantitative mapping of the cubic or tetragonal structures were impossible due to the nature of signal integration in TEM.

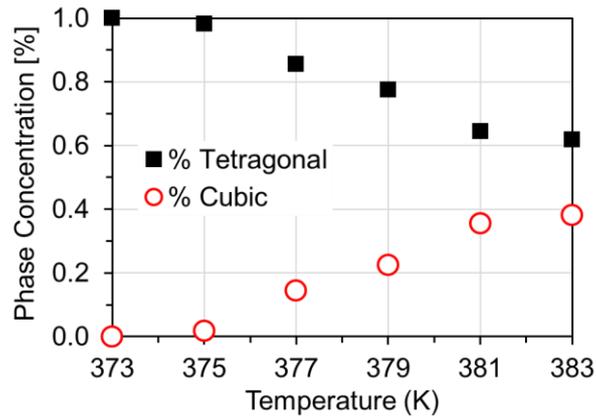

**Figure 4| Phase evolution around the Curie temperature in BaTiO$_3$.** Plotting the relative volume of the cubic phase as a function of temperature shows that during the nucleation state, 2% of the volume transferred to cubic 375 K and 14% of the unit cells changed their symmetry at 377 K. Above this temperature, more unit cells transferred to cubic during the growth process so that at 383 K already 40% of the material assumed a cubic structure.

## Discussions

Our data show that the ferroelectric-to-paraelectric transition comprises a clear distinction between nucleation and growth. The direct observations hence help us characterize the transition both qualitatively and quantitatively. First, we can discuss the size and location of the nuclei. Recent studies showed that the surface of BaTiO$_3$ crystals contains a high concentration of electro-chemo-mechanical defects,[35] while previous literature suggests that the surface demonstrates the preference of a cubic structure over tetragonal symmetry. It is therefore not surprising that the nucleation emerges near the surface (i.e. inhomogeneous nucleation), where the defects may contribute to lower energy penalty that is associated with the existence of small areas of a foreign phase within a long-range homogeneous crystal.[39]

Although a large number of theoretical works discuss the nucleus size and structure as well as nucleation-and-growth mechanisms in ferroelectrics, typically, the discussion framework covers domain switching under electric field and not temperature-driven ferroic transitions. As opposed to domain switching or domain formation under the electric field, the ferroelectric-to-paraelectric transition is accompanied by domain deletion or annihilation. Hence, it may still be valid to interpret partially our results by adopting an electric-field driven nucleation model. Shin et al. suggested[40] that for domain switching in a perovskite ferroelectric (PbTiO$_3$), the nucleus size is at the nanometer length and of a square-like shape. Our observations support this type of nucleation as opposed to much larger nucleus size of the large-angle pyramid or prolate spheroidal structures, e.g. as in Landauer's model for domain switching under an electric field for uniaxial ferroelectrics.[41] Yet, we believe that a dedicated model for the nucleation-and-growth process during the temperature-driven ferroic transition will contribute significantly to the understanding of the origin of ferroelectricity.

The importance of the nucleation process stems among the rest from the will to determine the driving force or energy that is required for the ferroic transition. These values are extracted from the driving force required for the existence of a small foreign region within a homogeneous structure:

$$\Delta U = -\Delta U_V + \Delta U_s \qquad (1)$$

as in the case of quantifying the minimum required activation electric field for domain switching,[40] where $\Delta U_V$ and $\Delta U_s$ are the nucleus volume and surface energies, respectively.

For a temperature-driven phase transition, a similar approach can be applied to calculate the driving energy of the transition. We can use Gibbs's free energy for Equation 1 and substitute $a^3 \Delta G_V$ for the volume energy as well as $\Delta G_s = 6a^2 \gamma$. Here, $a$ is the nucleus size, $\Delta u_V$ is the specific nucleation energy and $\gamma$ is the nucleus specific surface energy, which comprises the chemical energy between the nucleus and the hosting phase. Requiring minimum energy $\left(\frac{\partial \Delta G}{\partial a} = 0\right)$, we obtain the following dependence on the (minimal, $a^*$) nucleus size: $\Delta G_V = -\frac{4\gamma}{a^*}$. Substituting this dependence into Equation 1, and the nucleation energy is given by:

$$\Delta G^* = 2(a^*)^2 \gamma \qquad (2)$$

We can now use our observation of $a^*$ equals the length of two-unit cells (80 pm), calculate the nucleus specific surface energy $\gamma = 1.07$ eV per unit-cell area,[42] and extract the nucleation energy barrier: $\Delta G^* = 2.13$ eV.

The main contribution arrives likely from the mechanical and electric strain that is accompanied by the ferroelectric-to-paraelectric transition, especially thanks to the proximity to the high defect concentration near the $BaTiO_3$ surface. Indeed, the elastic energy associated with having an island of a (0.8 nm)×(0.8 nm)×(0.8 nm) box of cubic unit cells within a homogeneous tetragonal phase is rather close (4.0 eV, see SI).[43] The nucleation barrier calculated here is also comparable with the 2 eV calculated from the model of Shin et al.[40] for lead titanate, when bearing in mind 50-nm crystal size and using the activation-field values calculated by these authors for the highest reported temperature (300 K).

The electric energy may also play a certain role, but we believe it is much less influential, because the depolarization energy loss due to the emergence of a 0.8-nm cubic nucleus (0.06 eV) is a few orders of magnitude smaller than the mechanical energy contribution.[44,45]

The nucleation process is responsible for transferring 14.3% of the unit cells from tetragonal to cubic. Figure 4 shows that additional 26% of the volume was transitioning from ferroelectric to paraelectric by an inhomogeneous sidewalk growth from the surface towards the bulk when the temperature varied by only 6 K.

Such sidewalk growth is typically linked with surface roughness,[46] which is consistent with previous direct observations of BaTiO$_3$ crystalline surfaces at the atomic scale.[35,47] The directionality of the sidewalk growth in Figure 3 occurs along the [011] direction (note that careful attention should be put here, because TEM imaging does not allow us to distinguish properly between [011] and [001], while as explained below, we believe that [001] sidewalk is more plausible). Figure 2a shows that the cubic lattice parameter ($a_C$ = 4.0120 Å) is a better fit to the long lattice parameters of the tetragonal phase ($c_T$ = 4.0187 Å) than to the short lattice parameter ($a_T$ = 4.0086 Å). Hence, progression of the cubic phase along the *c* axis is accompanied by lower misfit energy than growth of the cubic phase towards the [100] or [010] directions.

As a final remark we would like to note that the above atomic-scale mapping allows us to describe the ferroic transition both qualitatively and quantitatively from the dipole-moment perspective. The analysis of these data enabled comparison between the thermally driven mechanism in this experimental study and atomistic models of ferroic domain switching.[40] However, we would like to encourage developments of dedicated atomistic models to describe the above observations as well as further experimental investigations of the phase-transition dynamics in ferroics and solid materials. Likewise, we would like to encourage experimental realization of the dipole-moment dynamics during electric-field switching in ferroelectric materials.

## Acknowledgement

The authors acknowledges support from the Zuckerman STEM Leadership Program, the Technion Russel Barry Nanoscience Institute, KLA-Tencor Israel as well as from the Israel Science Foundation (ISF) grant #1602/17. Likewise, we would like to thank Dr. Yaron Kauffman, Mr. Michael Kalina and Mr. Joshua Schechter for technical support, while we thank Prof. Yoed Tsur for supplying us with some of the BaTiO$_3$ crystals.

during the Cubic-Tetragonal Phase Transition in BaTiO3(001). *Appl. Phys. Lett.* **2018**, *113* (2), 022901 1-4.

# Supplementary Information for: "*Surface nucleation of the paraelectric phase in ferroelectric BaTiO$_3$: Atomic scale mapping*"

*Maya Barzilay,[1,2] Hemaprabha Elangovan[1,2] and Yachin Ivry[1,2,*]*

[1]Department of Materials Science and Engineering, Technion – Israel Institute of Technology, Haifa 3200003, Israel.

[2]Solid State Institute, Technion – Israel Institute of Technology, Haifa 3200003, Israel.

[*]Correspondence to: ivry@technion.ac.il.

## SI - Table of contents

# SI - Nucleation energy calculations

The direct observations of the dipole-moment dynamics around the ferroelectric-to-paraelectric transition in this work allowed us to quantify some of the relevant energy values that characterize these transitions. These values include the nucleation energy barrier and the mechanical and electrical contribution to this nucleation barrier. Moreover, these calculations were used to compare the nucleation process of the ferroic process to nucleation during domain switching under electric fields. Here we elaborate how these calculations were done.

- **Nucleation barrier calculation**

Nucleation energy barrier for a cubic nucleus is provided by the combination of volume free energy and the interfacial energy between the parent phase and the nucleus. Considering a cubic nucleus of length '$a$', total change in Gibbs free energy: $\Delta G = -a^3 \Delta G_V + 6a^2 \gamma$, where $\Delta G_V$ is the free volume energy of the nucleus and $\gamma$ is the interfacial surface energy.

At the nucleation energy barrier, the change in free energy becomes nil and is given as, $\frac{\partial(\Delta G)}{\partial a} = 0$, then the free energy equation becomes: $12a\gamma + 3a^2 \Delta G_V = 0$, which gives the expression for the critical nucleus size, $a^* = -\frac{4\gamma}{\Delta G_V}$. Feeding this expression in the free energy change expression provides the nucleation energy barrier, $\Delta G^* = 2(a^*)^2 \gamma$. Knowing the surface free energy for a TiO$_2$ terminated BaTiO$_3$ surface is 1.07 eV per unit cell and the critical nucleus size from our observations (Figure 3) as two unit cells, which is 0.8 nm, the nucleation energy barrier is: $\Delta G^* = 2.13$ eV (we assume that the observed nuclei are approximately the minimal size).

- **Nuclei strain energy calculation**

The elastic strain energy is given by: $S = \frac{1}{2} V E \epsilon^2$, where V is the volume of the nucleus, E is Young's modulus of BaTiO$_3$, which is for a 50-nm particle is 40 GPa,[43] and the strain $\epsilon$ is $\frac{c_T - a_C}{c_T}$. The lattice parameters of the tetragonal and cubic phase are $a_T$ = 4.0086 Å, $c_T$ = 4.0187 Å, and $a_C$ = 4.0120 Å, respectively ($c_T$ corresponding to 2θ = 45.08°, $a_T$ corresponding to 2θ = 45.2° in XRD profiling of 308 K and $a_C$ corresponding to 2θ = 45.16° in the XRD profiling 393 K. Both the XRD profiles are shown in Figure 2a). The obtained elastic strain energy is thus S = 1.67 eV. This value is in agreement earlier predictions of the nucleus barrier for electric-field domain switching.[40]

- **Nuclei depolarization energy calculation**

Depolarization energy ($U_d$) of BaTiO$_3$ was calculated based on Landauer's model:[41] $U_d = \frac{4\pi P_s^2}{\epsilon_0 \epsilon_C}$ V, where $P$ is the spontaneous polarization. For two-unit-cell nucleus of BaTiO$_3$, $P_s = 1.67 \times 10^{-7}$ C m$^{-2}$,[44] $\epsilon_0$ and $\epsilon_C$ are relative constants of the vacuum and the material (~1000)[45] respectively. Incorporating these values, $U_d$, is calculated to be 0.06 eV, which is much smaller than the value calculated for the mechanical contribution, which is 1.67 eV. The dominancy of mechanical contribution over electric contribution is in agreement with earlier predictions of the nucleus barrier for electric-field domain switching.[40]

## SI - Ion-displacement mapping

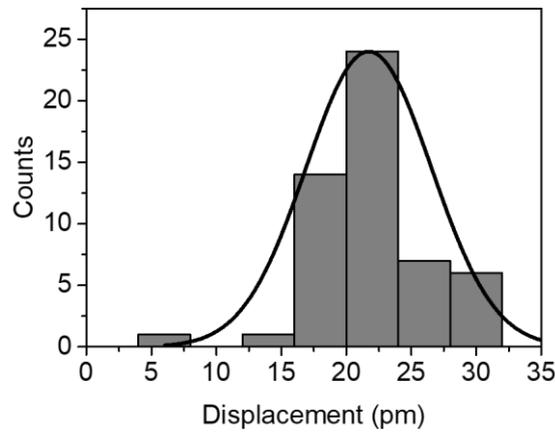

**Figure SI-1| Distribution of off-center Ti ion displacement in ferroelectric BaTiO$_3$.** Histogram of the off-center Ti ion displacements measured from Figure 2b, showing confidence dipole-moment mapping with 22±5 pm off-center Ti-ion displacement.

# SI - Large-scale TEM micrographs

The study of dipole-moment dynamics requires TEM imaging under variable conditions and over a longer duration than static domain studies require. Moreover, to avoid material damaging, exposure time to the electron beam must be minimized. Hence, tracking the dipole-moment dynamics in a specific area typically involves imaging of areas that slightly vary from one image to another, so that the area of interest is the regions that are overlapped between the different images. In addition, even if the region of interest is small, larger areas are typically imaged, again, to avoid damaging due to the irradiating electron beam.

To allow comparison between dipole-moment mappings at different times as well as to help emphasize the phase transition mechanism, the main text includes often closer look at the (unprocessed) TEM micrographs that help highlight the atomic structure. Here, in the Supplementary Information, we present the larger-scale micrograph as a reference.

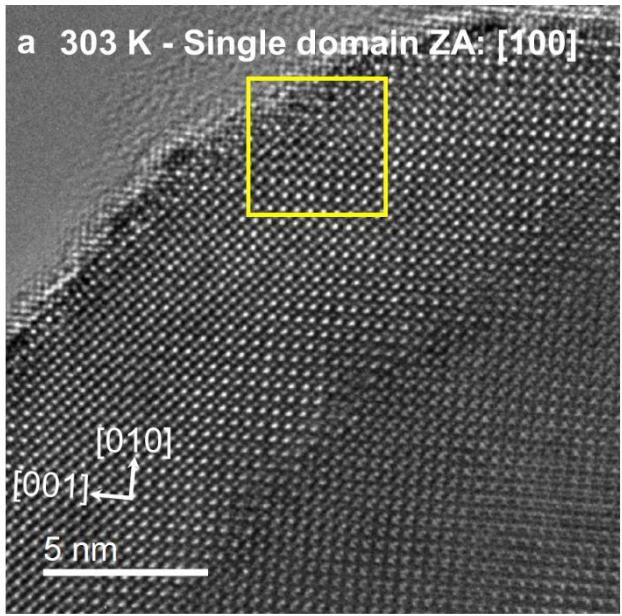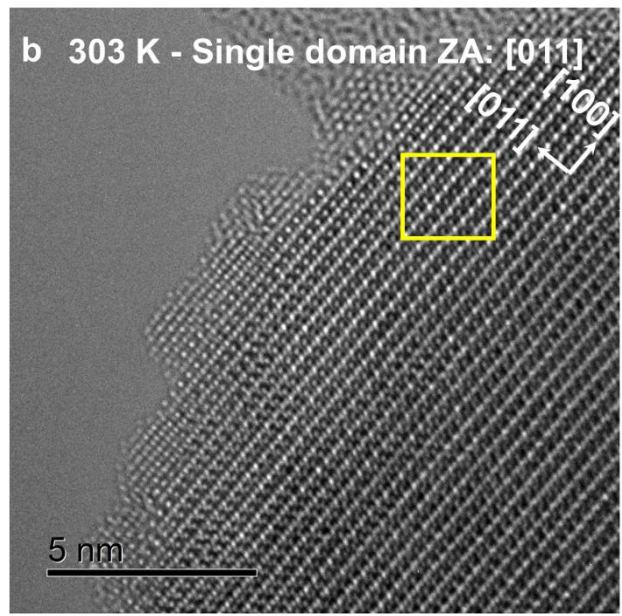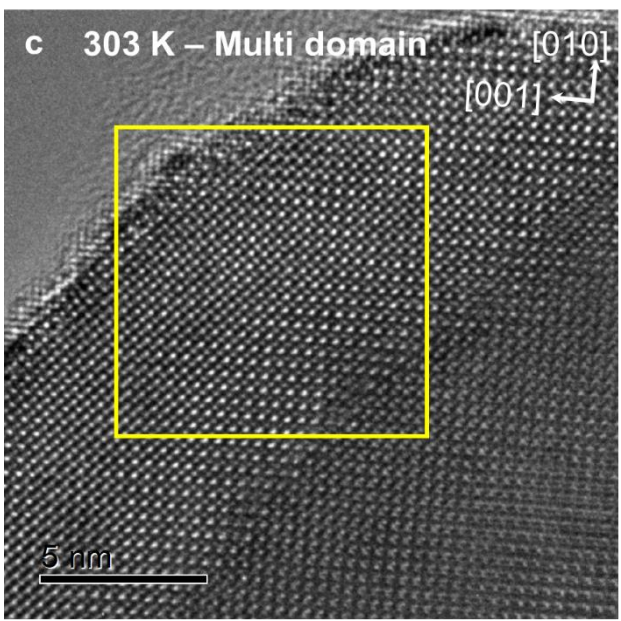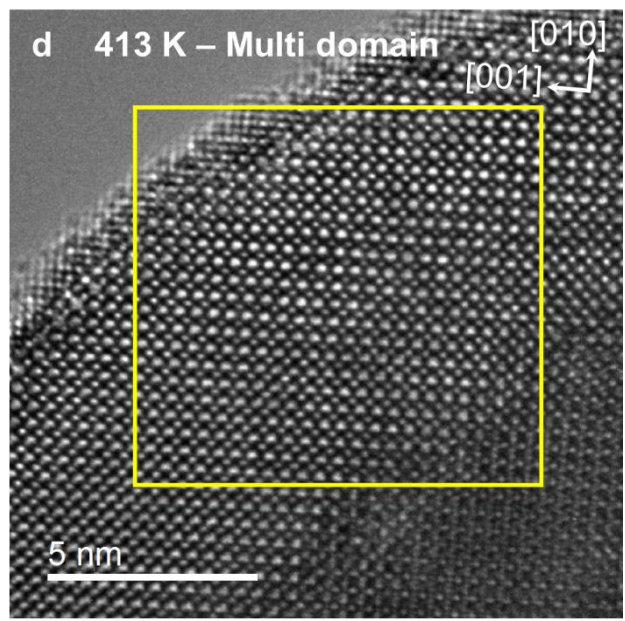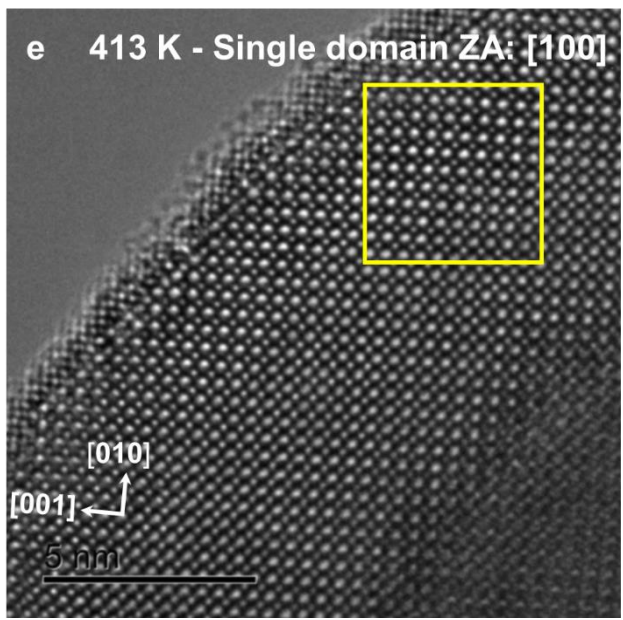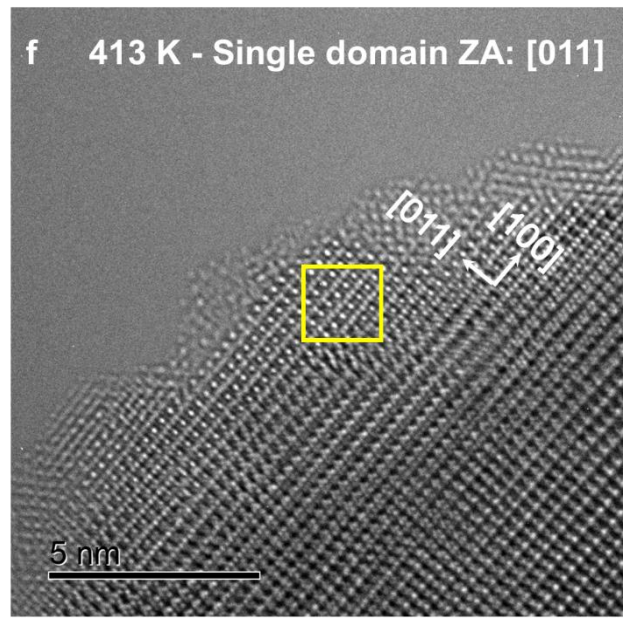

**Figure SI-2| Temperature-driven phase transition in BaTiO$_3$.** Large-scale TEM micrographs of BaTiO$_3$ nanoparticle at 303 K and 413 K at single domain and multi domains and at two different zone axes are given here. Here, (**a-b**), (**c-d**) and (**e-f**) correspond to Figures 2b-c, 2d-e and 2f-g, respectively.

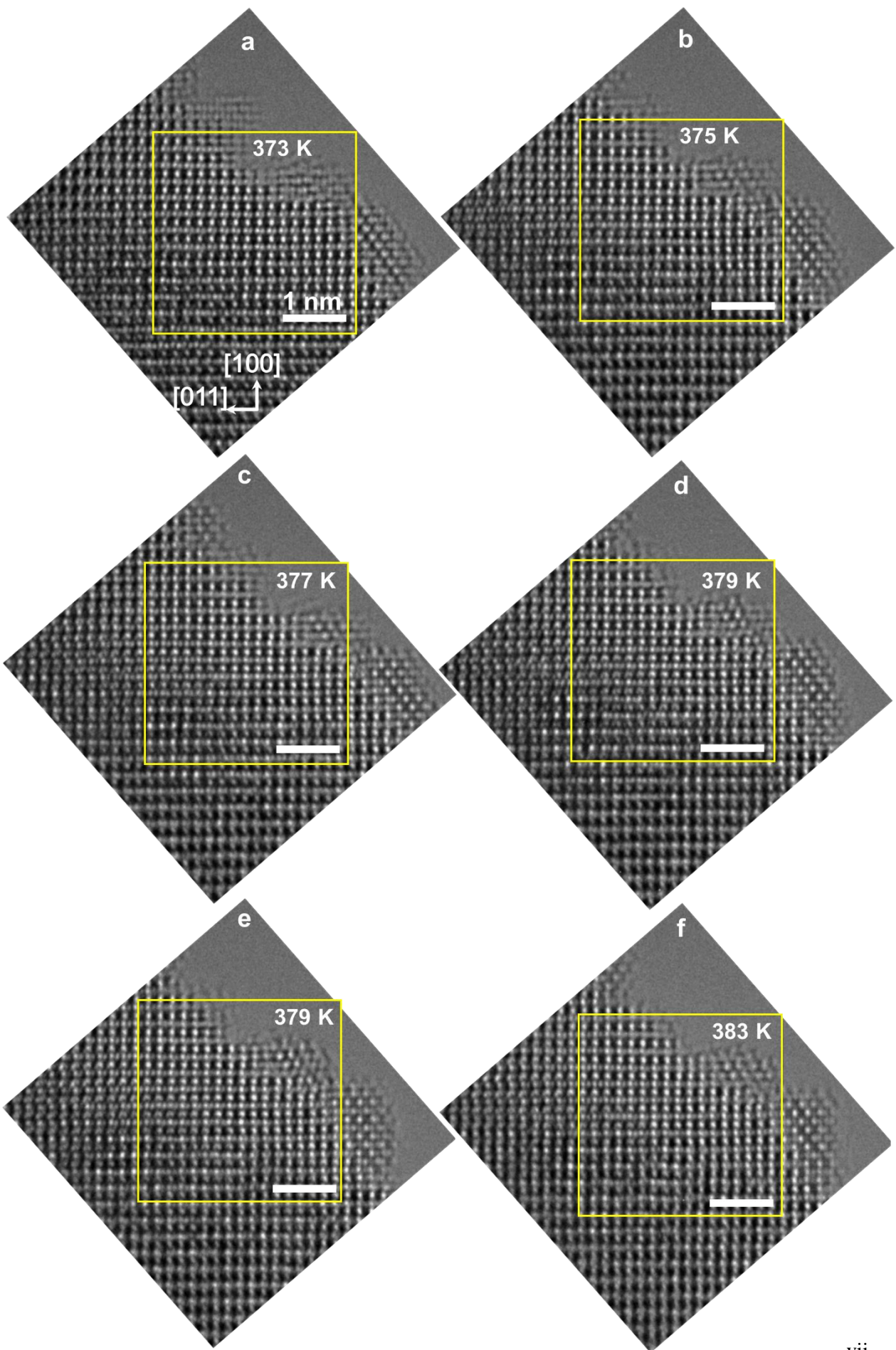

**Figure SI-3| Nucleation and growth of the paraelectric phase.** (**a-f**) Large-scale TEM micrographs of BaTiO$_3$ crystallites during in-situ heating from 373 K to 383 K. Highlighted area correspond to Figures 3a-f in the main text.